%% file: MnEuro.tex
\documentstyle[epsfig,amssym]{europhys}
\input euromacr

\begin{document}

\newcommand{\be}{\begin{equation}}
\newcommand{\ee}{\end{equation}}
\newcommand{\bea}{\begin{eqnarray}}
\newcommand{\eea}{\end{eqnarray}}

\euro{}{}{}{}
\Date{}
\shorttitle{M. N. Leuenberger {\it et al.} SPIN RELAXATION IN MN$_{12}$-ACETATE}

\title{Spin relaxation in Mn$_{12}$-acetate}

\author{Michael N. Leuenberger \And Daniel Loss}
\institute{Department of Physics and Astronomy, University of Basel \\
Klingelbergstrasse 82, 4056 Basel, Switzerland}

\rec{}{}

\pacs{
\Pacs{75}{45.+j}{Macroscopic quantum phenomena in magnetic systems}
\Pacs{75}{50.Tt}{Fine-particle systems}
\Pacs{75}{30.Pd}{Surface magnetism}
}

\maketitle

\begin{abstract}
We present a comprehensive theory of the magnetization relaxation in
a Mn$_{12}$-acetate crystal based on thermally assisted
spin tunneling induced by quartic anisotropy and weak transverse
magnetic
fields.
The overall relaxation rate as function of the magnetic field
is calculated and shown to agree well with data including all
resonance peaks. The Lorentzian shape of the resonances is also
in good agreement with recent data.
A generalized master equation including resonances is derived
and solved exactly. It is shown that many transition paths with
comparable weight exist that contribute to the relaxation process.
Previously unknown spin-phonon coupling constants are calculated
explicitly.
\end{abstract}

The molecular magnet Mn$_{12}$-acetate---a spin 10 system with large
easy-axis anisotropy---has attracted much recent interest\cite{Paulsen et al,Sessoli,Novak et al} since several experiments on the magnetization relaxation
have revealed pronounced
peaks\cite{Friedman,Thomas,Hernandez}  in response
to a varying magnetic field $H_z$ applied along the
easy axis of the  crystal.
These peaks occur  at field values where spin states become
pairwise degenerate.
Following
earlier suggestions\cite{Barbara,Novak2}, this  phenomenon has been
interpreted as a manifestation of resonant
tunneling of the magnetization.  However, in a  critical
comparison between  model
calculations\cite{Villain,Fort,Luis,VILLAIN,GARANIN,Gunther} and
experimental data\cite{Friedman,Thomas,FRIEDMAN} Friedman
{\it et al.}\cite{FRIEDMAN}
point out that a consistent explanation of the relaxation is still
missing. One of the main challenges for theory is to explain the
overall shape of the relaxation curve as well as the
Lorentzian shape of the measured resonance peaks\cite{FRIEDMAN}.

\begin{figure}[htb]
  \begin{center}
    \leavevmode
\epsfxsize=8cm
\epsffile{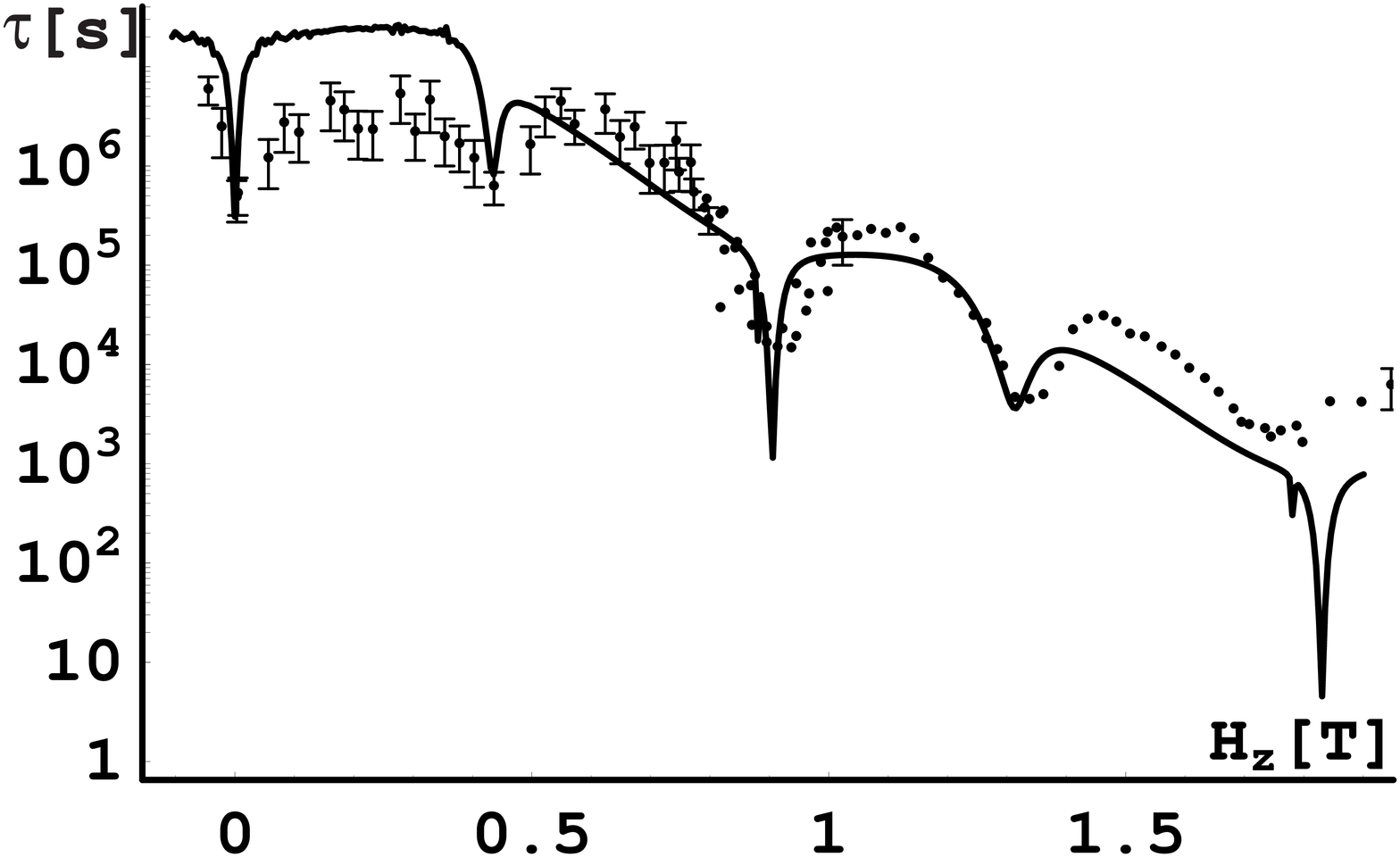}
  \end{center}
\caption{Full line: semilogarithmic plot of calculated relaxation time
$\tau$ as function of  magnetic field $H_z$ at $T=1.9$K. The optimal fit values
are:
$A/k_B=0.54$K, $B/k_B=1.1\cdot 10^{-3}$K,
and $B_4/k_B=4.3\cdot 10^{-5}$K, $\theta=1^\circ$, and
$c=1.45\cdot 10^3$m/s.
Dots and error bars: data\protect\cite{Thomas}.}
\label{relaxationtime}
\end{figure}

\begin{figure}
  \begin{center}
    \leavevmode
\epsfxsize=8cm
\epsffile{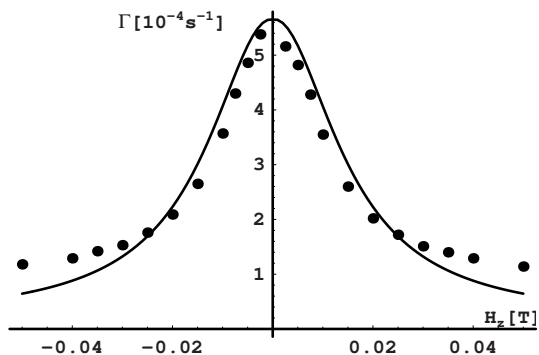}
  \end{center}
\caption{ Full line: plot of calculated relaxation rate
$\Gamma=1/\tau$ as function
of $H_z$ for the first resonance peak
at $T=2.6$K. The Lorentzian shape is due to $\Gamma_m^{m'}$ in
(\protect\ref{Lorentzian}). The optimal
fit values are: $A/k_B=0.56$K,
$B/k_B=1.3\cdot 10^{-3}$K, $B_4/k_B=10.2\cdot 10^{-5}$K,
{\bf $-1^\circ\le\theta\le 1^\circ$}, and  $c=2.0\cdot 10^3$m/s.
Dots: data\protect\cite{FRIEDMAN}. }
\label{singlepeak}
\end{figure}

\begin{figure}
  \begin{center}
    \leavevmode
\epsfxsize=8cm
\epsffile{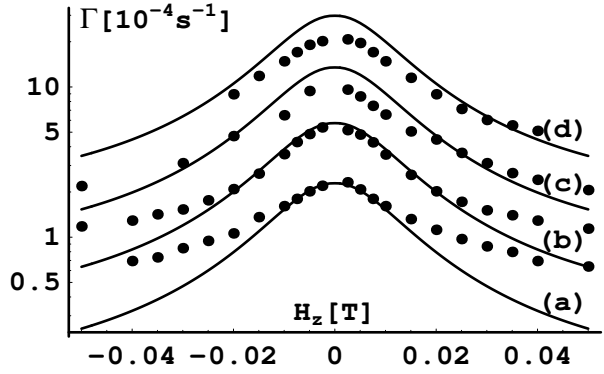}
  \end{center}
\caption{ Full lines: semilogarithmic plots of
calculated relaxation rate $\Gamma=1/\tau$
as function of $H_z$ for the first resonance peak
at (a) $T=2.5$K,
(b) $T=2.6$K, (c) $T=2.7$K, and (d) $T=2.8$K.
The optimal fit values are the same as in fig.2.
Dots: data\protect\cite{FRIEDMAN}. }
\label{threepeaks}
\end{figure}

In this work we shall present a
model calculation of the magnetization relaxation which is based
on thermally assisted tunneling.
We find for the first time reasonably good agreement both with the
overall relaxation rate (including the resonances) measured by Thomas {\it et
al.}\cite{Thomas} (see fig.\ref{relaxationtime})  and with the
Lorentzian shape of the
first resonance peaks (see figs.\ref{singlepeak},\ref{threepeaks})
measured
by Friedman {\it et al.}\cite{FRIEDMAN}
for four different temperatures.

The model introduced below contains five independent parameters:
three anisotropy constants $A\gg B\gg B_4$, a misalignment angle
$\theta$
(angle
between field direction and easy axis), and sound velocity $c$.
Previously unknown spin-phonon coupling constants are determined
for the first time; we find that, quite remarkably,  they can be expressed
in terms of $A$ only.
The constants $A,B,B_4$ are known within some
experimental uncertainty from
independent  measurements\cite{Barra,Zhong},
$\theta$ is typically about $ 1^\circ$\cite{FRIEDMAN}, while $c$ is not
measured
yet (to our knowledge).
We achieve optimal agreement between our theory and  data as follows.
In accordance with \cite{FRIEDMAN} we set
$\theta = 1^\circ$, while the
constants $A,B,B_4$ are fitted to the relaxation data by observing,
however,
the constraints that  $A,B,B_4$ are allowed to vary only in the range of
their  experimental uncertainties. The sound velocity $c$, on the
other hand, is a free fit parameter, for which we obtain
$c=(1.45-2.0)\cdot 10^3$ m/s, which is within expected range.
Thus, in contrast to previous
calculations\cite{Villain,Fort,Luis,VILLAIN,GARANIN,Fernandez} our theory
is in reasonably good agreement
not only with the relaxation data of refs.\cite{Thomas,FRIEDMAN} but with
all  experimental
parameter values known so far. In addition, a new prediction is made
(sound velocity $c$) which can be tested experimentally.

\begin{figure}
  \begin{center}
    \leavevmode
\epsfxsize=8cm
\epsffile{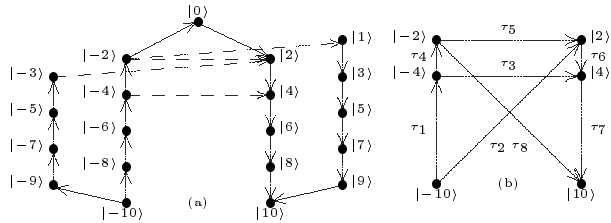}
  \end{center}
\caption{  (a) Spin relaxation paths (from $s=10$ to $s=-10$) for $0\leq g\mu_B H_z\leq A$. 
Dashed lines: tunneling transitions due to $B_4$- and $H_x$-terms. %
(b) Associated serially reduced diagram with relaxation times
$\tau_n$ given in (\protect\ref{analytic}). %
For $\left|H_z\right|\protect\lesssim 0.05$T only the path $\tau_1\rightarrow \tau_3\rightarrow\tau_7$ is relevant. }
\label{wiring}
\end{figure}

Extending previous
work\cite{Villain,Fort,Luis,VILLAIN,GARANIN,Fernandez} we
make use of a generalized master equation which treats phonon-induced
spin transitions
between nearest and next-nearest energy levels as well as  resonant
tunneling due to quartic anisotropies and transverse fields on the same
footing. We solve the derived master equation by exact diagonalization.
In addition, we identify for the first time the dominant
transition paths (see fig.\ref{wiring}) and show that  the
magnetization reversal is
not dominated by just one single term but rather by
several paths of comparable weight.
Finally we note that our model is minimal
in the sense
that it is sufficient to explain the existing data. Accordingly, we do
not include hyperfine fields\cite{Luis,Fernandez}, since the width of the resulting Gaussians\cite{Luis,VILLAIN} turns out to be smaller for $T\gtrsim 1$K than the width of the Lorentzians obtained below ({\it cf.} \cite{Wernsdorfer}).
Similarly, we  ignore dipolar\cite{Luis,Fernandez}
interactions  since
they have been ruled out by experiments on diluted
samples\cite{SESSOLI}.

{\it Model.} To describe the behavior of the Mn$_{12}$-acetate molecule
we follow earlier
work\cite{Villain,Fort,Luis,VILLAIN,GARANIN,HernandezHxTunneling}
and  adopt a single-spin Hamiltonian which includes spin-phonon
coupling.
Explicitly,
${\cal H}={\cal H}_{\rm a}+{\cal H}_{\rm Z}+{\cal H}_{\rm sp}+{\cal
H}_{\rm
T}$, where
\be
{\cal H}_{\rm a}=-AS_z^2-BS_z^4
\label{H_a}
\ee
describes the magnetic anisotropy with easy axis ($A\gg B> 0$)
along the z-direction. ${\bf S}$ is the
spin operator with $s=10$, and $A/k_B=0.52-0.56$
K\cite{Barra,Zhong}, and
$B/k_B=(1.1-1.3)\cdot 10^{-3}$ K\cite{Barra,Zhong}  are the
anisotropy
constants ($k_B$ is the Boltzmann constant). The Zeeman term through
which the
external magnetic field $H_z$ couples to the spin ${\bf S}$ is given by
${\cal H}_{\rm Z}=-g\mu_BH_zS_z$,
while the
tunneling between  $S_z$-states is described
by
\be
{\cal H}_{\rm T}=-\frac{1}{2}B_4\left(S_+^4+S_-^4\right)-g\mu_BH_x S_x\,
,
\label{H_T}
\ee
where  $H_x=|{\bf H}|\sin\theta$ ($\ll H_z$)
is the transverse field, with $\theta$
being the misalignment angle.
Here, $B_4/k_B=(4.3-14.4)\cdot 10^{-5}$ K\cite{Barra},
and $g=1.9$\cite{gSessoli}.
Finally, the most general spin-phonon coupling\cite{Callan} is described
by
\bea
{\cal H}_{\rm sp}& = & g_1(\epsilon_{xx}-\epsilon_{yy})\otimes
(S_x^2-S_y^2)+
\frac{1}{2}g_2\epsilon_{xy}\otimes\{S_x,S_y\} \\
& & +\left.\frac{1}{2}g_3(\epsilon_{xz}\otimes \{S_x,S_z\}+\epsilon_{yz}
\otimes\{S_y,S_z\})+\frac{1}{2}g_4(\omega_{xz}\otimes
\{S_x,S_z\}+\omega_{yz}\otimes\{S_y,S_z\})\, , \right. \nonumber
\label{H_sp}
\eea
where $g_i$ are the spin-phonon coupling constants, and
$\epsilon_{\alpha \beta} $ ($\omega_{\alpha\beta}$) is the
(anti-)symmetric part of the strain tensor defined through
the displacement ${\bf u}$ as $(\partial u_\alpha/\partial x_\beta \pm
\partial u_\beta/\partial x_\alpha)/2$.
To determine  $g_i$ occurring
in (\ref{H_sp}) we follow ref.\cite{Dohm} (full details will be given
elsewhere\cite{PRB}). From the displacement resulting from rotation
only,
${\bf u}=\delta\mbox{\boldmath $\phi$}\times{\bf x}$ (in leading order), we
obtain
the
infinitesimal
rotation angle $\delta\mbox{\boldmath $\phi$}=\frac{1}{2}{\rm
\nabla}\times{\bf u}=\left(\omega_{yz},\omega_{zx},
\omega_{xy}\right)$.
Rotating then the spin vector ${\bf S}$ we find (to
leading order in $\omega_{\alpha\beta}$) that the easy axis term,
$-AS_z^2$,
is transformed into
$A(\omega_{xz}\{S_x,S_z\}+\omega_{yz}\{S_y,S_z\})$. Comparison with
the last term in (\ref{H_sp}) then  yields $g_4=2A$.
Next, expanding the rotation matrices and ${\bf u}$
up to second-order, we find\cite{PRB},
$-\delta\phi_x^2= \varepsilon_{yy}+\varepsilon_{zz}-\varepsilon_{xx} $,
and cyclic.
After rotating the rhs of
$-AS_z^2=-A({\bf S}^2-S_x^2-S_y^2)$ we obtain a term of the form
$A\left(\epsilon_{xx}-\epsilon_{yy}\right)\left(S_x^2-S_y^2\right)$.
Comparing this with (\ref{H_sp}) we see
that $g_1=A$, and thus $g_1=g_4/2=A$.
Finally, we note that the terms  $\propto g_{1,2}$
produce second-order transitions, while the ones $\propto g_{3,4}$
produce
first-order transitions.  Thus,
it is very
plausible, in accordance with \cite{Abragam}, to adopt the
approximations,
$|g_2|\approx g_1=A$, and $|g_3| \approx g_4=2A$ (the sign is irrelevant
for the
transition rates calculated below).

We denote by
$\left|m\right>$, $-s\leq m\leq s$,
the eigenstate of the unperturbed Hamiltonian
${\cal H}_{\rm a}+{\cal H}_{\rm Z}$ with eigenvalue
$\varepsilon_{m}=-Am^2-Bm^4-g\mu_BH_zm$.
If the external magnetic field $H_z$ is increased one gets doubly
degenerate spin states whenever a level $m$ coincides with a level
$m'$ on the opposite side
of the well (separated by the barrier $\approx A$).
The resonance condition for double degeneracy,
{\it i.e.} $\varepsilon_{m}=\varepsilon_{m'}$,  leads to the resonance field
\be
{g\mu_B}H_z^{mm'}=-(m+m')\left[A+B\left(m^2+m'^2\right)\right].
\ee
As usual, we refer to $m+m'=$ even (odd) as even (odd) resonances.

{\it Master equation}. We describe the relaxation of the magnetization
in terms of a master equation for the reduced density matrix $\rho(t)$
which includes off-diagonal terms due to resonances.
Using the notation $\rho_{mm'}=\left<m\left|\rho\right|m'\right>$,
$\rho_m=\left<m\left|\rho\right|m\right>$,
we start from the generalized master equation\cite{Blum}
\bea
\dot{\rho}_{m'm}=\delta_{m'm}\sum_{n(\ne m)}
W_{mn}\rho_n-\gamma_{m'm}\rho_{m'm}\, ,
\label{GME}
\eea
where in the white-noise approximation
$\gamma_{m'm}=\gamma_{mm'}=(W_m +W_{m'})/2$, with
$W_m=\sum_{n(\ne m)}W_{nm}$\cite{PRB}.
The phonon induced transition rates $W_{mn}$ are
calculated below. Eq. (\ref{GME}) holds for correlation times $\tau_c$
much smaller than the spin relaxation time $\tau$
-- being satisfied here\cite{footnote1}.
Next we  incorporate resonant tunneling into (\ref{GME}).
Let
$\left|m\right>$ and
$\left|m'\right>$ be two  unperturbed eigenstates
on the lhs and rhs of the barrier, {\it resp.}, which are degenerate
when $\delta H_z=H_z^{mm'}-H_z$ vanishes, but get
detuned otherwise.
In the presence of tunneling, induced by ${{\cal H}}_T$, the two states
form
(anti-) symmetric levels split by $E_{mm'}$ (for $\delta H_z=0$).
$E_{mm'}$ is obtained by standard means\cite{Garanin}.
The 2-state  Hamiltonian including a bias field then reads
\be
{\overline{{\cal H}}}_T
=\xi_m\left|m\right>\left<m\right|
+\frac{E_{mm'}}{2}\left|m\right>\left<m'\right|
\,\,+ (m \leftrightarrow m'),
\ee
with $\xi_m=g\mu_B \delta H_z m$.
${\overline{{\cal H}}}_T$ provides a valid description as long as the
level
splitting
$\Delta=\sqrt{(\xi_m-\xi_{m'})^2+E_{mm'}^2}$ remains
smaller than
$|\varepsilon_{m}-\varepsilon_{m\pm 1}|$, and
$|\varepsilon_{m'}-\varepsilon_{m'\pm 1}|$.
Next we consider the quantum dynamics of the associated density matrix
adopting standard arguments\cite{Blum}. From the von Neumann equation
$\dot{\rho}=i\left[\rho,{\overline{{\cal H}}}_T\right]/\hbar$,
we get $\dot{\rho}_m  =
iE_{mm'}\left(\rho_{mm'}-\rho_{m'm}\right)/2{\hbar}$,
\bea
\dot{\rho}_{mm'}\!  =
\!-\!\left(\!\frac{i}{\hbar}\xi_{mm'}+\gamma_{mm'}\!\right)\!
\rho_{mm'}\!
+\frac{iE_{mm'}}{2\hbar}\! \left(\rho_m\!-\!\rho_{m'}\right),
\label{rho1}
\eea
where $\xi_{mm'}=\xi_{m}-\xi_{m'}$,
and likewise for $m\leftrightarrow m'$. Here we have allowed
for damping due to phonons ({\it i.e.} finite life-time of the levels $m,m'$)
according to (\ref{GME}).
Inserting the stationary solution of (\ref{rho1})
into the $\dot{\rho}_m$--equation,
we get
$\dot{\rho}_m  =  \Gamma_m^{m'}\left(\rho_{m'}-\rho_m\right)$,
where
\be
\Gamma_m^{m'}=E_{mm'}^2
\frac{W_m+W_{m'}}{\xi_{mm'}^2
+\hbar^2\left(W_m+W_{m'}\right)^2/4 }
\label{Lorentzian}
\ee
is the transition rate from $m$ to $m'$ (induced by tunneling)
in the presence of
phonon-damping. We note that  $\Gamma_m^{m'}$ has a Lorentzian shape
with respect to the external
magnetic field $\delta H_z$ (the $H_z$-dependence of $W_m$
around the resonances turns
out to be much weaker); it is this $\Gamma_m^{m'}$ that will determine
the
peak shape of the magnetization resonances (see below
and figs.\ref{relaxationtime}-\ref{threepeaks})\cite{Lorentzian}.

To obtain the complete master equation we include also
the  transitions involving non-resonant levels and get
\be
\dot{\rho}_m  = -W_m\rho_m+\!\!\!\sum_{n\ne m,m'}\!\!\!
W_{mn}\rho_n\,+\Gamma_m^{m'}(\rho_{m'}-\rho_m),
\label{finalmeq}
\ee
and the same with $m\leftrightarrow m'$ (see ref.\cite{PRB} for details).
A similar equation has been used in ref.\cite{Fort}, whose $\Gamma_m^{m'}$
has been obtained through a
different and rather lengthy derivation \cite{Villainetal}.
For levels $k\neq m,m'$, eq. (\ref{finalmeq}) reduces to
$\dot{\rho}_k=-W_k\rho_k+\sum_nW_{kn}\rho_n$.

As in ref.\cite{VILLAIN} we evaluate the spin-phonon rates
via Fermi golden rule (with thermal averaging over phonons)
and find\cite{PRB}
\bea
W_{m\pm 1,m} & = & \frac{A^2s_{\pm 1}}{12\pi\rho c^5\hbar^4}
\frac{(\varepsilon_{m\pm 1}-\varepsilon_m)^3}
{e^{\beta(\varepsilon_{m\pm 1}-\varepsilon_m)}-1}\, , \\
W_{m\pm 2,m} & = & \frac{17A^2s_{\pm 2}}{192\pi\rho c^5\hbar^4}
\frac{(\varepsilon_{m\pm 2}-\varepsilon_m)^3}%
{e^{\beta(\varepsilon_{m\pm 2}-\varepsilon_m)}-1}\, ,
\eea
where
$s_{\pm
1}=(s\mp m)(s\pm m+1)(2m\pm 1)^2$, and
$s_{\pm 2}=(s\mp m)(s\pm m+1)(s\mp m-1)(s\pm m+2)$.
The mass density $\rho$ is given by $1.83\cdot 10^3$kg/m$^3$\cite{Lis}.
Here, $c$ is the sound velocity of the Mn-crystal,
which is
the only free parameter in our theory. It has not been considered before and  we are not aware of any
measurements of c.

{\it Relaxation time}.
We solve now the master equation
exactly to find the largest ({\it i.e.} dominant) relaxation time.
To this end it is convenient to rewrite (\ref{finalmeq})
as $\dot{\vec {\rho}}(t)={\tilde W} {\vec \rho}(t)$,  where $\rho_n$
are the components of ${\vec {\rho}}$.
The rate matrix ${\tilde W}$ has 21 eigenvalues $w_i$ found
by exact diagonalization. The eigenvalues $w_i$ turn out to be
non-degenerate
with the smallest one being  far separated from the remaining ones.
The relaxation time is then obtained from  $\tau=\max_i\left\{-1/{\rm
Re}\,{w_i}\right\}$.
The result is plotted in fig.\ref{relaxationtime}, where
the overall relaxation rate $\tau$ is shown as function of $H_z$ and at
$T=1.9$K.
We
note that in our model the even resonances are induced by the quartic
$B_4$-anisotropy, whereas the odd resonances are induced by
product-combinations
of $B_4S_\pm^4$- and $H_x S_x$-terms.
In accordance with the experimental
uncertainty\cite{FRIEDMAN} we keep $-1^\circ\le\theta\le 1^\circ$ for all plots. Then the relaxation
time $\tau$ at an even resonance peak is about 3 times smaller than the
one at a subsequent odd resonance peak. The relevant tunnel
splitting energies of the even and odd  resonances are about the same:
$E_{4,-4}\approx E_{2,-3}\approx E_{3,-5}\approx E_{1,-4}\approx k_B\cdot
20$mK. ($E_{2,-2}/k_B\approx 0.5$K.) For comparison we also include in
fig.\ref{relaxationtime} the data reported by Thomas {\it et
al.}\cite{Thomas,angle}. We have optimized the  fit  (as explained in the
introduction) in such a way that the fits of the model parameters, given
by
$A/k_B=0.54$K, $B/k_B=1.1\cdot 10^{-3}$K,
and $B_4/k_B=4.3\cdot 10^{-5}$K, $\theta=1^\circ$, are roughly within the
reported experimental uncertainties\cite{Barra,Zhong}(see above),
while the fit of the sound velocity yields $c=1.45\cdot 10^3$m/s.
Almost identical plots are obtained for $0.5^\circ\lesssim\theta\lesssim
3^\circ$ \cite{PRB,footnoteHX}.

In figs.\ref{singlepeak},\ref{threepeaks} we plot the peaks of the
first resonance at $H_z=0$  (induced only by the $B_4$-anisotropy)
for four different temperatures, $T=2.5,2.6,2.7,2.8$K. The peaks (like all
others) are of single Lorentzian shape as a result of the 2-state transition
rate $\Gamma_m^{m'}$ given in (\ref{Lorentzian}) (see also \cite{Lorentzian}).
For comparison we plot in figs.\ref{singlepeak}, \ref{threepeaks} the
data
reported by Friedman {\it et al.}\cite{FRIEDMAN} for the same temperatures
(no error bars, however, are given).
The optimal fit values are: $A/k_B=0.56$K,
$B/k_B=1.3\cdot 10^{-3}$K, $B_4/k_B=10.2\cdot 10^{-5}$K, for all $-1^\circ\le\theta\le 1^\circ$,
and $c=2.0\cdot 10^3$m/s.
Note that these values are the same for all four temperatures.
These fitting parameters turn out to be somewhat
larger than in fig.\ref{relaxationtime},
which could be caused by differences {\it e.g.} in
volume-to-surface ratio and/or in shape anisotropy, etc. Indeed,
the sample of ref.\cite{FRIEDMAN} consists of many small crystallites in
contrast
to the single crystal used in ref.\cite{Thomas}.
In any case, the differences are small, and the sound velocity $c$
seems to be within the expected order of magnitude.
Clearly, it would be highly desirable to check this prediction by an
independent measurement of $c$.
To sum up,
the agreement between theory and experiment is reasonably good, and
our model and its evaluation contains the essential physics responsible for the
magnetization relaxation.

{\it Relaxation paths}. It is instructive to determine the dominant
transition paths via which the spin can relax into its ground state.
For this we derive an approximate analytic expression for the relaxation
time denoted by $\tau^*$ (to distinguish it from the exact $\tau$
obtained above).
First, we solve the master equation for one {\it particular} transition
path $n$ which does not intersect with other paths.
For $H_z\ge 0$ we find (details will be given elsewhere\cite{PRB}) :
\be
\tau_n=\frac{1}{1+e^{\beta(\varepsilon_s-\varepsilon_{-s})}}
\sum_{\{m\}_n}\frac{e^{\beta(\varepsilon_{m}-\varepsilon_{-s})}}
{\gamma_m},
\label{analytic}
\ee
where $\gamma_m =W_{mm'}$, or $\Gamma_m^{m'}$, depending
on the particular path n characterized by the sum over the levels $m$
(see fig.\ref{wiring}).
Eq. (\ref{analytic}) holds for arbitrary initial ($\varepsilon_i$)
and final ($\varepsilon_f$) energies; for $H_z=0$
({\it i.e.} $\varepsilon_i=\varepsilon_f$)
eq. (\ref{analytic}) reduces to the result found
by Villain {\it et al.}\cite{VILLAIN}. If there is more than one dominant
path (which is typically the case) we have to account for intersections
at vertices.
For this we
associate with each path a probability current $J_n=\dot{\rho}_n$,
and interprete  eq. (\ref{analytic}) as a serial circuit
with the summands playing the role of ``resistances".
This  allows us then
to set up  flow diagrams for $J_n$ (see fig.\ref{wiring}),
which obey the analog of Kirchhoff's rules:
(K1) $\sum_n J_n=0$: The sum over all incoming and outgoing currents
vanishes at a vertex (current conservation).
(K2) $\sum_n J_n\tau_n=\Delta N$: The sum over all voltage drops
($J_n\tau_n$)
is equal to the source-drain voltage  $\Delta N=\rho_s-\rho_{-s}$
for any closed path (probability conservation).
The total probability current is given by $J={\Delta {\dot N}}$.
Fig.\ref{wiring} shows the  complete (a) and its serially reduced
(b) flow diagram for $ 0\leq H_z\le A/{g\mu_B}$.
{}From (K1) we get $J=J_1+J_2$, $J_2+J_5=J_6$, $J_1=J_3+J_4$,
$J_3+J_6=J_7$, $J_4=J_5+J_8$, and $J_7+J_8=J$,
while from (K2) we get
$\Delta N=J_1\tau_1+J_3\tau_3+J_7\tau_7$,
$J_3\tau_3=J_4\tau_4+J_5\tau_5+J_6\tau_6$,
$J_2\tau_2=J_1\tau_1+J_4\tau_4+J_5\tau_5$,
$J_8\tau_8=J_5\tau_5+J_6\tau_6+J_7\tau_7$.
From these equations
we obtain the relaxation time defined by $\tau^*=\Delta N/J$\cite{PRB}.
Finally, when plotted as function of $H_z$ there is no visible
difference
between the exact $\tau$ and this approximate $\tau^*$, which confirms
that
the diagram in fig.\ref{wiring} contains
the physically relevant relaxation paths close to the first resonance.
Similar results are obtained for the other resonances\cite{PRB}.


We are grateful to A. Chiolero and T. Pohjola for  useful discussions. This work has been supported  by the Swiss NSF.

{\it Additional Remark.}
After submission of this work, we were notified by J.R. Friedman that our prediction of the sound velocity $c$ is in very good agreement with recent specific heat measurements\cite{Gomes} (see \cite{PRB} for details).

\end{document}

%% file: euromacr.tex
\def\And{{\rm and\ }}

\newif\ifboo \boofalse

\def\Review#1{\boofalse{\it #1},}
\def\Name#1{{\sc #1},}
\def\Vol#1{\ifboo Vol. {\bf #1}\else{\bf #1}\fi}
\def\Year#1{\ifboo #1\else(#1)\fi}

\def\Page#1{\ifboo {\rm p. #1}\else{\rm #1}\fi}

%% file: MnEuro.bbl
\begin{thebibliography}{99}
\bibitem{Paulsen et al} \Name{C. Paulsen {\it et al.}}
\Review{J. Magn. Magn. Mater.} \Vol{140-144} \Year{1995} \Page{379}.
\bibitem{Sessoli} \Name{R. Sessoli {\it et al.}}
\Review{Nature (London)} \Vol{365} \Year{1993} \Page{141}.
\bibitem{Novak et al} \Name{M.A. Novak {\it et al.}}
\Review{J. Magn. Magn. Mater.} \Vol{146} \Year{1995} \Page{211}.
\bibitem{Friedman} \Name{J.R. Friedman {\it et al.}}
\Review{Phys. Rev. Lett.} \Vol{76} \Year{1996} \Page{3830}.
\bibitem{Thomas} \Name{L. Thomas {\it et al.}}
\Review{Letters to Nature} \Vol{383} \Year{1996} \Page{145}.
\bibitem{Hernandez} \Name{J.M. Hern\'andez {\it et al.}}
\Review{Europhys. Lett.} \Vol{35} \Year{1996} \Page{301}.
\bibitem{Barbara} \Name{B. Barbara {\it et al.}}
\Review{J. Magn. Magn. Mater.} \Vol{140-144} \Year{1995} \Page{1825}.
\bibitem{Novak2} \Name{M.A. Novak, R. Sessoli} in
\Review{Quantum tunneling of  Magnetization} eds. L. Gunther, B. Barbara
(Kluwer, Dordrecht, 1995).
\bibitem{Villain} \Name{J. Villain {\it et al.}}
\Review{Europhys. Lett.} \Vol{27} \Year{1994} \Page{159}.
\bibitem{Fort} \Name{A. Fort {\it et al.}}
\Review{Phys. Rev. Lett.} \Vol{80} \Year{1998} \Page{612}.
\bibitem{Luis} \Name{F. Luis {\it et al.}}
\Review{Phys. Rev. B} \Vol{57} \Year{1998} \Page{505}.
\bibitem{VILLAIN} \Name{F. Hartmann-Boutron {\it et al.}}
\Review{Mod. Phys. B} \Vol{10} \Year{1996} \Page{2577}.
\bibitem{GARANIN} \Name{D.A. Garanin, E.M. Chudnovsky} \Review{Phys. Rev. B} 
\Vol{56} \Year{1997} \Page{11102}.
\bibitem{Gunther} \Name{L. Gunther, Europhys. Lett.} \Vol{39} \Year{1997} \Page{1}.
\bibitem{FRIEDMAN} \Name{J.R. Friedman {\it et al.}}
\Review{Phys. Rev. B} \Vol{58} \Year{1998} \Page{R14729}.
\bibitem{Barra} \Name{A.L. Barra {\it et al.}}
\Review{Phys. Rev. B} \Vol{56} \Year{1997} \Page{8192}.
\bibitem{Zhong} \Name{Y. Zhong {\it et al.}}
condmat/9809133.
\bibitem{Fernandez} \Name{F. Fern\'andez {\it et al.}} \Review{J. Appl. Phys.} \Vol{83} \Year{1998} \Page{6940}.
\bibitem{Wernsdorfer} \Name{W. Wernsdorfer {\it et al.}} preprint.
\bibitem{SESSOLI} \Name{R. Sessoli} \Review{Mol. Cryst. Liq. Cryst.} \Vol{274} \Year{1995} \Page{145}.
\bibitem{HernandezHxTunneling} \Name{J.M. Hern\'andez {\it et al.}} \Review{Phys. Rev. B}
\Vol{55} \Year{1997} \Page{5858}.
\bibitem{gSessoli}
\Name{R. Sessoli {\it et al.}}
\Review{J. Am. Chem. Soc.} \Vol{115} \Year{1993} \Page{1804}.
\bibitem{Callan} \Name{E. Callan, H. Callan} \Review{Phys. Rev.} \Vol{139A} \Year{1965} \Page{455}.
\bibitem{Dohm} \Name{V. Dohm, P. Fulde} \Review{Z. Phys. B} \Vol{21} \Year{1975} \Page{369}.
\bibitem{PRB} \Name{M.N. Leuenberger, D. Loss} to be published.
\bibitem{Abragam} \Name{A. Abragam, A. Bleany} \Review{Electron Paramagnetic
Resonance of Transition Ions} Clarendon Press, Oxford, 1970.
\bibitem{Blum} \Name{K. Blum} \Review{Density Matrix Theory and Applications, 2nd
edition} (Plenum Press, 1996), ch. 8.
\bibitem{footnote1} An estimate for thermal phonons yields
$\tau_c\sim \hbar/k_B T\sim 10^{-11}$ s, at $T=1$K,
whereas $\tau\sim 1$ s.
\bibitem{Garanin} \Name{D.A. Garanin} \Review{J. Phys. A} \Vol{24} \Year{1991} \Page{L61}.
\bibitem{Lorentzian} Note that in the $\tau(H_z)$ plot these Lorentzians are truncated by the spin-phonon transition rates $W_m$ and $W_{m'}$ in such a way that the effective linewidth is much larger than $(W_m+W_{m'})/2$\cite{PRB}.
\bibitem{Villainetal} \Name{J. Villain {\it et al.}} \Review{J. Phys. I France} \Vol{7} \Year{1997} \Page{1583}.
\bibitem{Lis} \Name{T. Lis} \Review{Acta Crystallogr. Sec. B} \Vol{36} \Year{1980} \Page{2042}.
\bibitem{footnoteHX}
The present theory holds for  $\left | H_x\right |\lesssim 1000$ G (which
is well satisfied here); otherwise the shift of the  levels $\left |m\right>$
due to $H_x S_x$ must be taken into account\protect\cite{PRB}.
\bibitem{angle} In \cite{Thomas} the correct temperature is $T=1.9$K, not 2.1K (B. Barbara, W. Wernsdorfer, priv. comm.).
\bibitem{Gomes} \Name{A.M. Gomes {\it et al.}}
\Review{Phys. Rev. B} \Vol{57} \Year{1998} \Page{5021}.

\end{thebibliography}
